\documentclass[letterpaper, 10 pt, journal, twoside]{IEEEtran}

\usepackage{graphicx}
\usepackage{subfigure}
\usepackage{lipsum}
\usepackage{cite}

\usepackage{amsfonts} %
\usepackage{amsmath} %
\usepackage{amssymb}
\usepackage{physics}

\usepackage{fancyhdr}
\usepackage{color}
\usepackage[lined,boxed,commentsnumbered,ruled]{algorithm2e}
\usepackage{url}

\usepackage{booktabs}
\usepackage[table,x11names,dvipsnames,table]{xcolor}
\usepackage{colortbl}
\graphicspath{
    {./fig/}
}
\usepackage{diagbox}
\usepackage[normalem]{ulem}
\usepackage{float}

\usepackage{wrapfig}

\newcommand{\revision}[2]{{#2}}
\newcommand{\srevision}[1]{{#1}}

\ifCLASSINFOpdf
\else
\fi

\begin{document}
\title{Learning to Accelerate Decomposition for Multi-Directional 3D Printing}

\author{Chenming Wu$^{1}$, Yong-Jin Liu$^{1}$ and Charlie C.L. Wang$^{2}$ %
\thanks{This work was partially supported by the GRF grant from the Research Grants Council of the Hong Kong SAR, China (Project No.: CUHK/14202219). \emph{(Corresponding Author: Yong-Jin Liu and Charlie C.L. Wang.)} } 
\thanks{$^{1}$C. Wu, Y.-J. Liu are with the Department of Computer Science and Technology, Tsinghua University, Beijing 100084, China (email: wcm15@mails.tsinghua.edu.cn; liuyongjin@tsinghua.edu.cn).}%
\thanks{$^{2}$C.C.L. Wang is with the Department of Mechanical and Automation Engineering, The Chinese University of Hong Kong, Shatin, Hong Kong (email: cwang@mae.cuhk.edu.hk).}%
\thanks{Digital Object Identifier (DOI): see top of this page.}
}

\markboth{}
{Wu \MakeLowercase{\textit{et al.}}: Learning to Accelerate Decomposition for Multi-Directional 3D Printing}

\maketitle

\makeatletter
\newcommand{\removelatexerror}{\let\@latex@error\@gobble}
\makeatother

\begin{abstract}
Multi-directional 3D printing has the capability of decreasing or eliminating the need for support structures. Recent work proposed a beam-guided search algorithm to find an optimized sequence of plane-clipping, which gives volume decomposition of a given 3D model. Different printing directions are employed in different regions to fabricate a model with tremendously less support (or even no support in many cases). To obtain optimized decomposition, a large beam width needs to be used in the search algorithm, 
leading to a very time-consuming computation. In this paper, we propose a learning framework that can accelerate the beam-guided search by using a smaller number of the original beam width to obtain results with similar quality. Specifically, we use the results of beam-guided search with large beam width to train a scoring function for candidate clipping planes based on six newly proposed feature metrics. With the help of these feature metrics, both the current and the sequence-dependent information are captured by the neural network to score candidates of clipping. As a result, we can achieve around $3 \times$ computational speed. We test and demonstrate our accelerated decomposition on a large dataset of models for 3D printing.
\end{abstract}

\begin{IEEEkeywords}
Additive Manufacturing, Intelligent and Flexible Manufacturing
\end{IEEEkeywords}

\IEEEpeerreviewmaketitle

\section{Introduction}
\label{sec:intro}
\IEEEPARstart{3}{D} printing makes the rapid fabrication of complex objects possible. \revision{Due to i}{I}ts capability has been developed in many scenarios -- from the micro-scale fabrication of bio-structures to \revision{on}{in}-situ construction of architecture. However, Fused Deposition Modeling~(FDM) using planar layers with fixed 3D printing direction suffers from the need of support structures (shortly called \textit{support} in the following context), which are used to prevent the collapse of material in overhang regions due to gravity. 
Supports bring in many problems, including hard-to-remove, surface damage and material waste as summarized in~\cite{Hu2015a}.%

\begin{figure}[t]
\centering
\includegraphics[width=\linewidth]{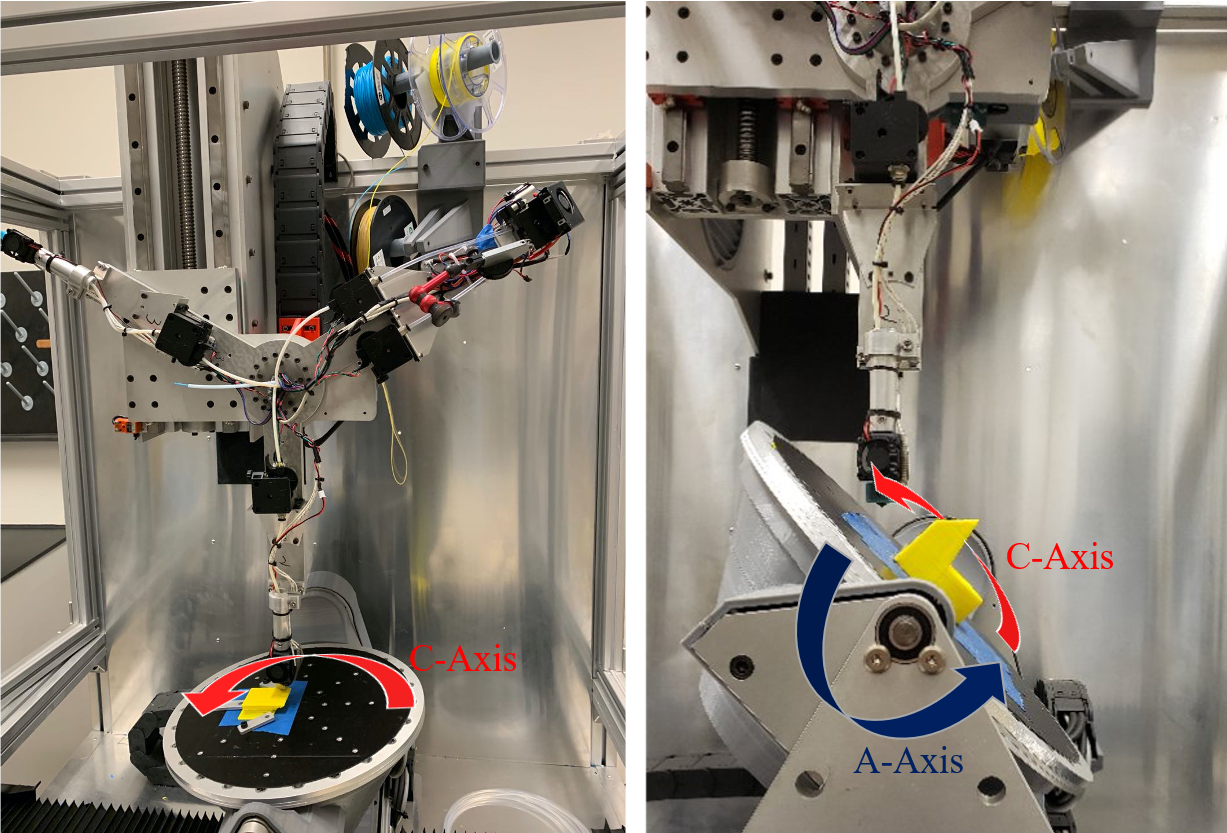}
\caption{A 5-DOF multi-directional 3D printing system that can deposit material along \srevision{with} different directions: (left) the printer head can move along $x-$, $y-$ and $z-$axes and (right) the working table can rotate around two axes (see the arrows for the illustration of A-axis and C-axis).}
\label{fig:Hardware}
\end{figure}

\begin{figure}[!t]
\centering
\includegraphics[width=\linewidth]{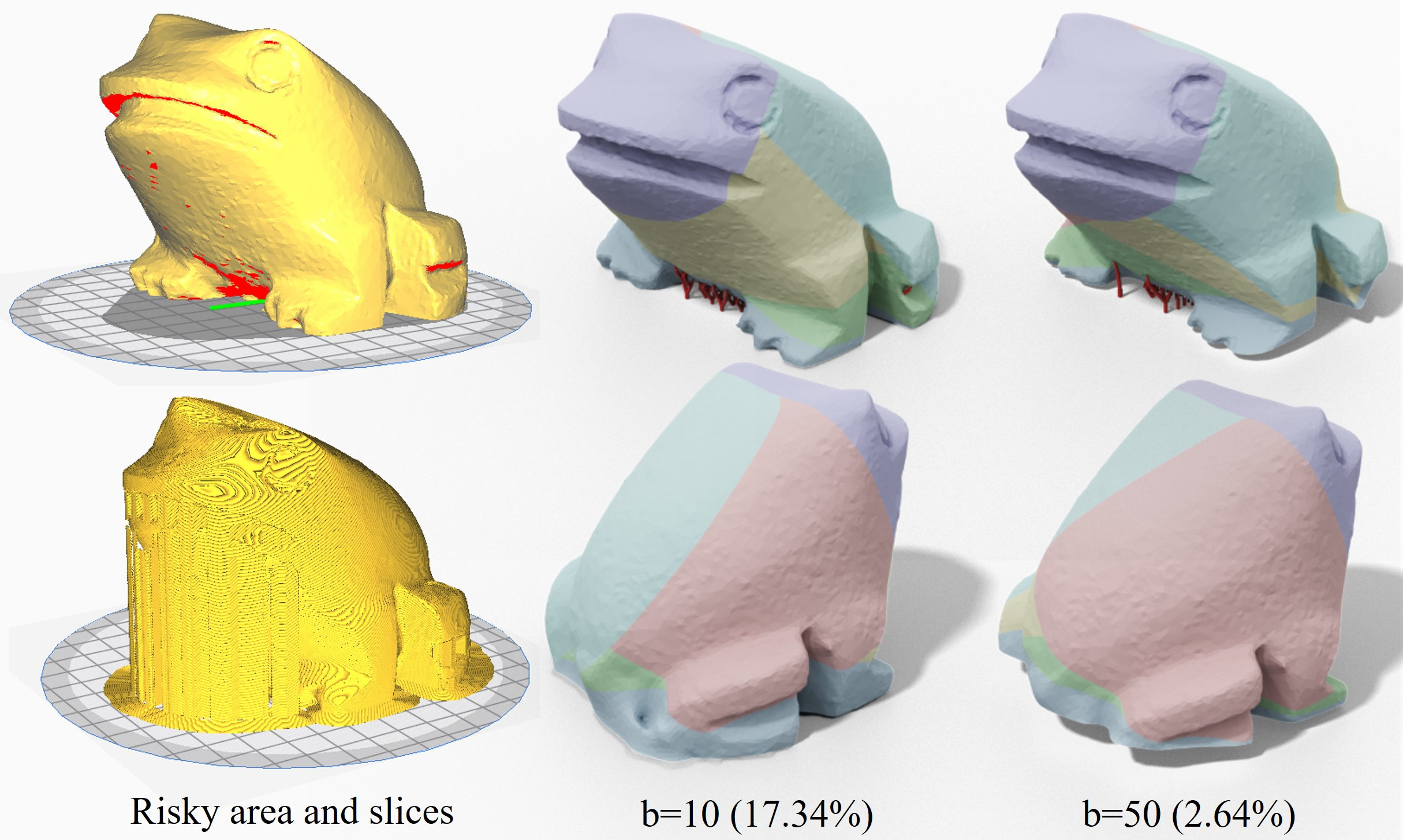}
\caption{
\revision{}{An example of using different widths in beam search is given on the frog model (\textit{ID: 81368} from the \textit{Thingi10k} dataset~\cite{zhou2016thingi10k}). \revision{Large amount}{A large number} of supports are needed by using the conventional 3D printing (left). Multi-directional 3D printing can significantly reduce the need of supports, and the regions need additional supports can be reduced from $17.34\%$ (middle) to $2.64\%$ (right) of the total area when the beam width increases from $10$ to $50$. Regions to be printed along different directions are displayed in different colors to represent the results of volume decomposition, and supporting structures represented by red struts are added.}
}
\label{figDifferentB}
\end{figure}

To avoid using a large number of supports, our previous work~\cite{wu2019general} proposes an algorithm to decompose 3D models into a sequence of sub-components, the volume of which can be printed one by one along \srevision{with} different directions for different components. Candidates clipping planes are used as a set of samples to define the search space for determining an optimized sequence of decomposition. Different criteria are defined to ensure the feasibility and the manufacturability (e.g., collision-free and no floating region, etc.). The most important part of the work \srevision{presented in} \cite{wu2019general} is a beam-guided search algorithm with progressive relaxation. The benefit of the \revision{proposed}{beam} search algorithm is that it can avoid being stuck in local minimum \revision{compared to greedily searching the best result}{-- a common problem of greedy search}. Beam width $b=10$ is empirically used to balance the trade-off between computational efficiency and searching effectiveness. Though conducting a parallel implementation running on a computer with Intel(R) Core(TM) i7 CPU (4 cores), the method still results in an average computing time of 6 minutes.
On the other hand, using $b=10$ is \revision{only for the sack of the trading-off}{a compromise between performance and efficiency}. \revision{Obviously,}{Using} a larger $b$ would give us better results since the search space is expanded linearly when $b$ increases -- see Fig.\ref{figDifferentB} for an example. 

One question is, can we learn from the results generated by a large beam width so that even a search using a small beam width can produce comparable results?
Our answer is \textit{yes}. To achieve this goal, we propose to learn a \revision{classifier for a pair of}{scoring function for} candidate clipping planes by using \revision{five newly proposed}{six} feature metrics. With the help of these feature metrics, both the current and the sequence-dependent information are captured by the \revision{classifier}{neural network} to score candidates \revision{of}{for} clipping. The learning is conducted on the results of beam-guided search with large beam width (i.e., $b=50$) running on a large dataset of models for 3D printing, \textit{Thingi10k}, recently published by~\cite{zhou2016thingi10k}. As a result, we can achieve \revision{2 times}{around 3 times} acceleration while still keeping the similar quality on the results of volume decomposition.

In summary, we make the following contributions:
\begin{itemize}
    \item A learning-to-accelerate framework that can rank a set of candidate planes that best-fit the optimal results sampled on the large dataset, \srevision{which significantly accelerates the beam search algorithm without sacrificing the performance}.
    
    \item A method to convert the trajectories generated during \srevision{the} beam-guided search to \revision{pairwise comparisons}{listwise ranking orders at distinct stages} for training.
\end{itemize}
The computational efficiency of the proposed work is much better than our previous work~\cite{wu2019general} while keeping the quality of searching result\srevision{s} at \revision{the}{a} similar level.
The implementation of learning-based acceleration presented in this paper, together with the solid decomposition approach presented in \cite{wu2019general} is available at GitHub\footnote{\url{https://github.com/chenming-wu/pymdp/}}.

\section{Related Work}
\label{sec:relatedwork}
The problems caused by support have motivated a lot of research effort\srevision{s} to reduce the need for supports. There are three significant threads of research towards this goal: 1) proposing better patterns of supports so that the number of supports is smaller than the one generated by \revision{vanilla}{} support generators (ref.~\cite{vanek2014clever, dumas2014bridging}); 
2) segmenting digital models into several pieces, each of which can be built in a support-free or support-effective manner; 
3) using high degree-of-freedom (DOFs) robotic systems to automatically change the build direction so that the overhanging regions become safe regions that can be safely fabricated without \srevision{the need of }{}supports. Here we mainly review the prior work in the last two threads that are most relevant.

\subsection{Segmentation-based Methods}
A digital model can be first segmented into different components for fabrication and then assembled back to form the original model.
There are \revision{a number of}{several} methods that have explored to use segmentation to reduce the need of supports.
Hu et al. \cite{Hu2014} invented an algorithm to automatically decompose a 3D model into parts in approximately pyramidal shapes to be printed without support. Herholz et al. \cite{Herholz2015} proposed another algorithm to solve a similar problem by enabling slight deformation during decomposition where each component\revision{s}{} is in the shape of height-fields. RevoMaker \cite{Gao2015UIST} fabricated digital models by 3D printing on top of an existing cubic component, which can rotate itself to fabricate the shape of height-fields. Wei et al. \cite{wei18supportfree} partitioned a shell model into a small number of support-free parts using a skeleton-based algorithm. Muntoni et al. \cite{muntoni2018heightblock} also tackled the problem of decomposing a 3D model into a small set of non-overlapped height field blocks, which can be fabricated by either molding or AM. These methods are mostly algorithmic systems that can be easily incorporated into off-the-shelf manufacturing devices. However, the capability of manufacturing hardware has not been considered in the design of algorithms. 

\subsection{Multi-directional and Multi-axis Fabrication}
\revision{R}{The r}ecent development in robotic systems enable\srevision{s} researchers to think about a more flexible AM routine~\cite{urhal2019robot}. Adding more DOFs into the process of 3D printing seems promising and has gained \revision{a lot of}{much} attention. 
Keating and Oxman \cite{Keating2013} proposed to use a 6-DOF manufacturing platform driven by a robotic arm to fabricate the model either in an additive or subtractive manner. Pan et al.~\cite{Pan2014} rethink the process of \revision{}{\textit{Computer Numerical Control}} (CNC) machining and proposed a 5-axis motion system to accumulate materials. 
\textit{On-the-Fly Print} system proposed by Peng et al. \cite{peng2016fly} is a fast, interactive printing system modified from an off-the-shelf Delta printing device but with two additional DOFs. Based on the same system, Wu et al. \cite{wu2016printing} proposed an algorithm that can plan the collision-free printing orders of edges for wireframe models. 

Industrial robotic arms have been widely used in AM. For example, 
Huang et al. \cite{huang2016framefab} built up a robotic system for 3D printing wireframe models on a 6-DOF KUKA robotic arm. 
Dai et al. \cite{dai2018support} developed a voxel-growing algorithm for support-free printing digital models using a 6-DOF UR robotic arm. 
Shembekar et al.~\cite{shembekar2018trajectory} proposed a method to fabricate conformal surfaces by collision-free 3D printing trajectories on a 6-DOF robotic arm. 
To reduce the expense of hardware, a $3+2$-axis additive manufacturing is also proposed recently~\cite{xu18supportfree}. They adopted a flooding algorithm to plan \revision{a }{}collision-free and support-free paths. However, this approach can only be applied to tree-like 3D models with simple topology. Volume decomposition\srevision{-}based algorithms have been proposed in our prior work (ref.~\cite{wu2017RoboFDM,wu2019general}).

\begin{figure*}[t]
\centering
\includegraphics[width=\linewidth]{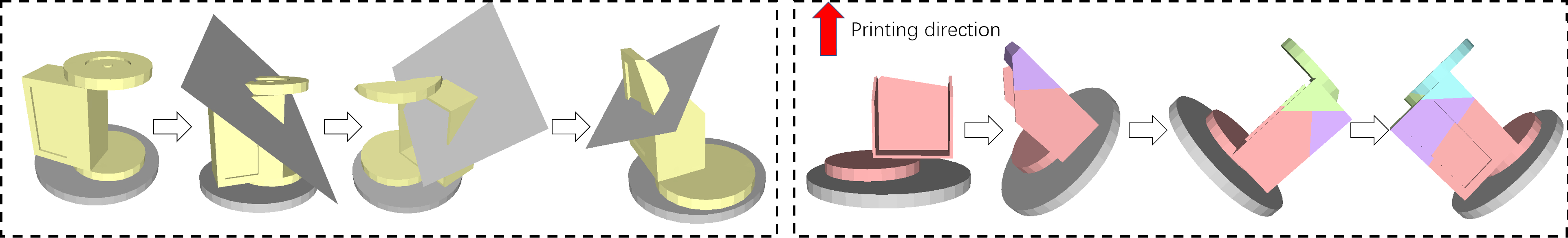}
\caption{A sequence of multi-directional 3D printing can be determined by computing a sequence of planar clipping (left), where the inverse order of clipping gives the sequence of multi-directional 3D printing (right). Details can be found in \cite{wu2019general}.}
\label{fig:OverviewClipping}
\end{figure*}

\subsection{Learning to Accelerate Search}

\revision{Searching feasible solution is a common problem in computer science, where most problems with huge search space are difficult to tackle. It is crucial to develop an efficient algorithm to solve search problems. The methodology of incorporating machine learning to accelerate search has been proved to be useful. For example, optimizing a program using different predefined operators is a combinatorial problem, and it is difficult to search for an optimal solution. The work of Chen et al.~\cite{chen2018learning} learned domain-specific statistical cost models to guide the search of tensor implementations over many possible choices for efficient deep-learning deployments. Recently, Adams et al.~\cite{adams2019learning} improved a beam search algorithm for Halide program optimization. They proposed to learn a cost model to predict runtime that takes the input of the derived features. Similarly, our work also aims at optimizing sequences of operations because applying different cuts in different stages also formulates an ample search space, and different orders matter.
A simple but effective method is proposed in this paper.}{
Efficiently searching a feasible solution is a common problem in computer science, where most problems have ample search space and thus challenging to tackle. Recent research advances the state-of-the-art by incorporating machine learning techniques. For example, optimizing a program using different predefined operators is a combinatorial problem that is difficult to optimize. The work of Chen et al. ~\cite{chen2018learning} learned domain-specific models with statistical costs to guide the search of tensor implementations over many possible choices for efficient deep-learning deployments. Recently, Adams et al.~\cite{adams2019learning} improved a beam search algorithm for Halide program optimization. They proposed to learn a cost model to predict running time by using the input of the derived features. We aim at searching optimal sequences of operations as applying different cuts in different stages. Learning a scoring function is similar to the problem solved in \cite{adams2019learning}.}

\srevision{Direct establishing a mapping from features to a score by supervised learning is difficult. Differently, we adopt the learning-to-rank (LTR)~\cite{liu2009learning} technique to solve our problem. LTR is one of the traditional topics in information retrieval, which aims at ranking a set of documents given a query (e.g., a keyword) by a user. LTR learns a scoring model from query-document features, thereafter the predicted scores can be used to order (rank) the documents. There are three types of LTR approaches: pointwise, pairwise, and listwise. Pointwise LTR approach learns a direct mapping from a single feature to an exact score~\cite{liu2009learning}. Pairwise LTR approach learns pairwise information between two features and convert the pairwise relationships to a ranking~\cite{burges2005learning, burges2010ranknet}. Listwise LTR approach treats a permutation of features as a basic unit and learns the best permutation~\cite{cao2007learning,guiver2009bayesian,xia2008listwise}. Our work is motivated by the idea of ranking the query-document's features using listwise LTR. A scoring function with our model-plane features as input is learned to accelerate the beam search procedure.}

\section{Preliminaries and Denotations}
\label{sec:preliminaries}
This section briefly introduces the idea of the beam-guided algorithm previously proposed in~\cite{wu2019general}.

\subsection{Problem Formulation}
Whether fabricating a model $\mathcal{M}$ layer-by-layer needs additional supports can be determined by if \textit{risky faces} exist on the surface of $\mathcal{M}$. A commonly used definition of identifying a risky face $f$ is 
\begin{equation}
e(f, \pi)=
\begin{cases}
1 & \mathbf{n}_f \cdot \mathbf{d}_{\pi} + \sin{( \alpha_{max} ) > 0}, \\
0 & \text{otherwise}.
\end{cases}
\end{equation}
where $\mathbf{d}_{\pi}$ (as the normal of $\pi$) gives the printing direction defined by a base plane $\pi$, $\mathbf{n}_f$ is the normal of $f$ and $\alpha_{max}$ is the maximal self-supporting angle (ref.\cite{Hu2015a}). Face $f$ is risky if $e(f, \pi)=1$ and otherwise it is called \textit{safe}.

In~\cite{wu2019general}, a multi-directional 3D printer is supervised by fabricating a sequence of parts decomposed from $\mathcal{M}$ where:
\begin{itemize}
\item $N$ components decomposed from $\mathcal{M}$ satisfies
\begin{equation}
\mathcal{M} = \mathcal{M}_1 \cup \mathcal{M}_2 \cup \cdots \cup \mathcal{M}_N = \cup_{i=1}^{N} \mathcal{M}_i
\end{equation}
with $\cup$ denoting the \textit{union} operator;

\item $\{ \mathcal{M}_{i=1,\cdots,N} \}$ is an ordered sequence that can be collision-freely fabricated with \begin{equation}
\pi_{i+1} = \mathcal{M}_{i+1} \cap \left( \cup_{j=1}^{i} \mathcal{M}_j \right) 
\end{equation}
being the base plane of $\mathcal{M}_{i+1}$, where $\cap$ denotes the \textit{intersection} operator;

\item $\pi_1$ is the working platform of a 3D printer;

\item All faces on a sub-region $\mathcal{M}_i$ are \textit{safe} according to $\mathbf{d}_{\pi_i}$ determined by $\pi_i$.
\end{itemize}
To achieve the decomposition satisfying all the above requirements, we use planes $\pi$ to cut $\mathcal{M}$.
If every clipped sub-region satisfies the manufacturability criteria, we could use the inverse order of clipping as the sequence of printing for the multi-directional 3D printers (see Fig.\ref{fig:OverviewClipping} for an illustration). The printing direction of a sub-part $\mathcal{M}_i$ is determined by the normal of the clipping plane.

We formulate the problem of reducing the area of risky faces on $\mathcal{M}_i$ as a problem that minimizes
\begin{equation}\label{eqGlobalObjective}
J = \sum_{i} \sum_{f \in \mathcal{M}_i} e(f, \pi_i) A(f)
\end{equation}
where $A(f)$ is the area of \srevision{a} face $f$\srevision{. As we are handling models represented by triangle meshes, the computation of $A(f)$ is straightforward. The metric $J$ is employed to measure the quality of different sequences of volume decomposition}. While minimizing the objective function defined in Eq.(\ref{eqGlobalObjective}), we need to ensure the manufacturability of each component.

\begin{figure}[t]
\centering
\includegraphics[width=\linewidth]{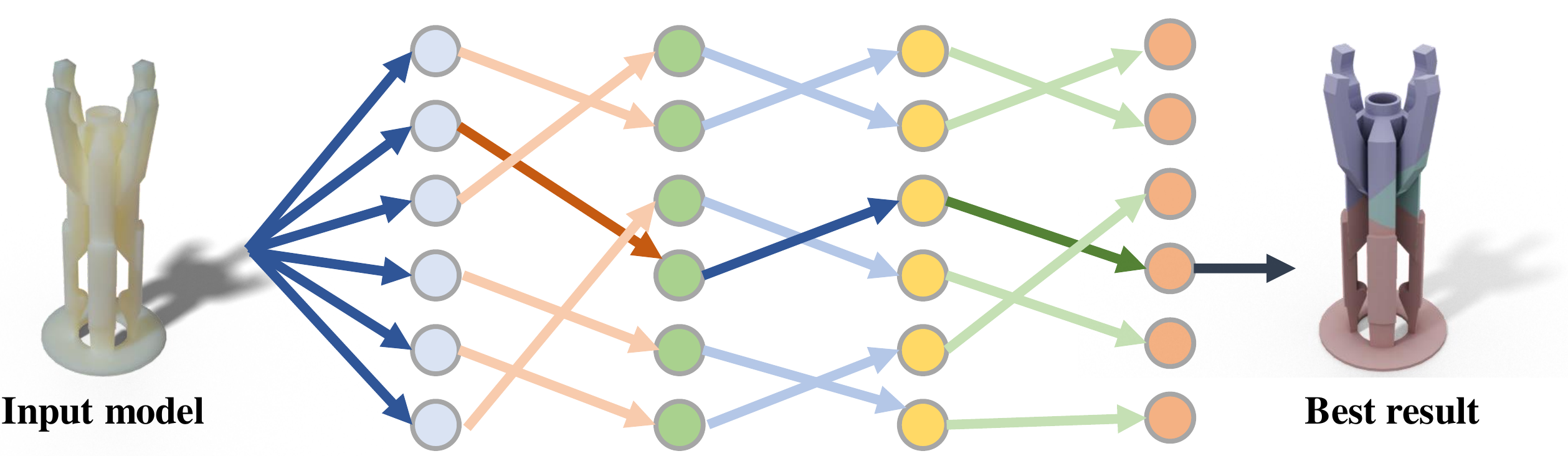}
\caption{An example of beam-guided searching trajectories generated in a case with $b=6$. The trajectory in dark color is the best trajectory $\tau^*$ giving the lowest value of $J$. The trajectories shown in light colors are the other trajectories having smaller values of $J$ than the one of $\tau^*$.}
\label{figTrajectory}
\end{figure}

\begin{figure*}[t]
\centering
\includegraphics[width=\linewidth]{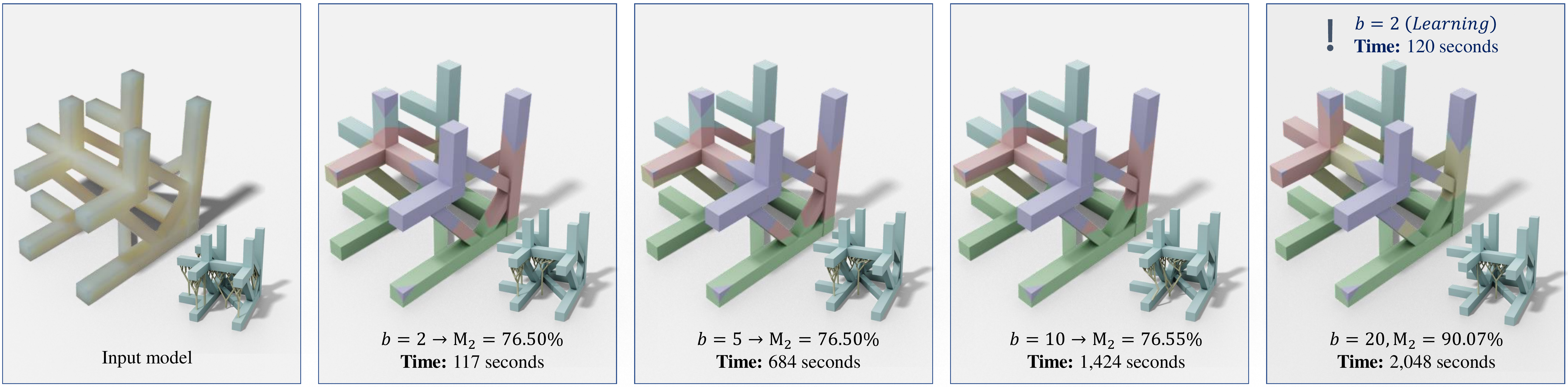}
\caption{An example in the \textit{Thingi10k} dataset (\textit{ID: 109926}). Our learning-based method outperforms the beam-guided search algorithms with small beam width of $b$. Here, $M_2$ indicates the percentage of \srevision{the} reduced risky area when using multi-directional 3D printing -- the higher the better. 
\srevision{From left to right, the results of \srevision{the} conventional beam search \cite{wu2019general} when using different widths. It can be observed that better results with less risky area can be obtained when using large beam width. With the help of the scoring function $G(\cdot)$ learned in this paper, we can use a very small beam width (i.e., $b=2$) to obtain the same result obtained by large beam width (i.e., $b=20$) in conventional beam search -- see the result shown in the right-most. Support structures are generated for multi-directional 3D printing by the method presented in \cite{wu2019general} and given in the bottom-right corner of each column, where the less risky area results in less amount of support.}
}\label{figTeaser}
\end{figure*}

\subsection{Beam-guided Search}
The beam-guided search is to optimize Eq.(\ref{eqGlobalObjective}). Considering the manufacturing constraints as well as search efficiency, we define four constraints in beam-guided search.

\vspace{2pt} \noindent \textbf{Criterion I:} All faces on $\mathcal{M}_i$ should be self-supported. 

\vspace{2pt} \noindent \textbf{Criterion II:} The remained model obtained by every clipping should be connected to the printing platform $\mathcal{P}$.

\vspace{2pt} \noindent \textbf{Criterion III:} The physical platform of the printer $\mathcal{P}$ is always below the clipping plane.

\vspace{2pt} \noindent \textbf{Criterion IV:} 
It is always preferred to have a large solid obtained above a clipping plane so that a large volume of solid can be fabricated along one fixed direction.
\vspace{5pt}

A beam-guided search~\cite{Lowerre1976} algorithm is proposed to guide the search. It builds a search tree that explores the search space by expanding promising nodes ($b$ nodes as \textit{beam width}) instead of the best one greedily (see Fig.\ref{figTrajectory}). It integrates the restrictive Criterion I (and its weak form) as an objective function to ensure that the beam search is broad enough to include both the local optimum and configurations that may lead to a global optimum. The other three criteria should also be satisfied during the beam-guided search. Defining the residual risky area of a model $\mathcal{M}_k$ according to a clipping plane \revision{$\gamma$}{$\pi$} as
\begin{equation}
    R(\mathcal{M}_k,\pi)= \sum_{f \in \mathcal{M}^+_k} e(f, \pi)A(f),
\end{equation}
where \revision{$\gamma$}{$\pi$} separates $\mathcal{M}_k$ into the part $\mathcal{M}^+_k$ above \revision{$\gamma$}{$\pi$} and the part $\mathcal{M}^-_k$ below \revision{$\gamma$}{$\pi$}. The proposed beam-guided search algorithm starts from an empty beam with the most restrictive requirement of $R(\mathcal{M}_{k},\pi) < \delta$, where $\delta$ is a threshold progressively increasing from a tiny number (e.g., 0.0001). Candidate clipping planes that satisfy this requirement and remove larger areas of risky faces have \srevision{a} higher priority to fill the $b$ beams. If there are still empty beams after the first `round' of filling, we relax $\delta$ by letting $\delta=5\delta$ until all $b$ beams are filled. Detail algorithm can be found in \cite{wu2019general}.

\section{Learning to Accelerate Decomposition}
\label{sec:algorithm}
\subsection{Methodology}
The beam-guided algorithm ~\cite{wu2019general} constrains the search space by imposing the manufacturing constraints (Criteria II \& III) and the volume heuristic (Criterion IV) while progressively relaxing the selection of `best' candidates (Criterion I). Larger beam width $b$ keeps more less-optimal candidates, which will \revision{then }{}have better chance \revision{to obtain}{of obtaining} a globally optimal solution. We conduct an experiment on the Thingi10k dataset to compare different choices of $b$, and it turns out that the average performance by $b=50$ is around $17\%$ better than the average performance generated by $b=1$ while it takes more than $36\times$ computing time to obtain those results. One example is given in the right of Fig.\ref{figTeaser}. 
The experimental results encourage us to explore the feasibility of learning from the underlying experience produced by large beam width of $B$, and utilizing the learned policy to guide a more effective search, which only keeps a much smaller \srevision{value of} beam width $b$ ($b\ll B$) during the search procedure.

Specifically, given $b$ nodes for configurations kept in the beam, we will be able to obtain thousands of candidates \revision{of}{for} the next cut. The original method presented in \cite{wu2019general} is employed to select the `best' and the relaxed `best' $B$ candidates ($b \ll B$). Here we will not keep all these $B$ candidates in the beam. Instead, only $b$ candidates are selected from these $B$ candidates, where the selection is conducted with the help of a \revision{pair-comparison based classifier}{scoring function} $G(\cdot)$ using six enriched feature metrics as input for each candidate clipping. An illustration \revision{for}{of} this selection procedure can be found in Fig.\ref{figPipeline}. The \revision{classifier}{scoring function} is constructed by \revision{decision-tree based ensemble}{a neural network}, which is trained by using the samples \revision{learned }{}from conducting beam-guided searches \cite{wu2019general} on \textit{Thingi10k} -- a large dataset of 3D printing models with a large beam width $B=50$.

In the rest of this section, we will first provide the enriched feature metrics. Then, we present the details of the accelerated search algorithm and the method to generate training samples. Lastly, the learning model of \revision{classifier}{the scoring function} is introduced.

\subsection{Featurization of Candidate Clipping}\label{subsec:Metrics}
We featurize each candidate cut to a vector consisting of six metrics. The metrics are carefully designed according to the criteria given in Sec.~\ref{sec:preliminaries}, which consider both the current and the sequence-dependent information for the configuration of a planar clipping. Note that, it is \revision{very important}{crucial} to have metrics to cover the sequence-dependent information (i.e., $M_2$ and $M_6$ below). Otherwise, it has \srevision{a} trivial chance to learn the strategy of beam-guided search that will not be stuck at local optimum when using large beam width. 

\vspace{5pt} \noindent \textbf{Ratio of reduced risky area $M_1$:} The reduced risky area is essentially the decreased value of Eq.(\ref{eqGlobalObjective}). $M_1$ is defined as the ratio of decreased risk area caused by a candidate clipping plane (the \textit{candidate}) over the value of $J$. 

\vspace{5pt}\noindent \textbf{Accumulated ratio of reduced risky area $M_2$:} Different stages have different values of $M_1$, which only reflect a local configuration. We define the sum of $M_1$ as the accumulated ratio of reduced risky area\srevision{s} to describe the situation of a sequence of planning. In short, $M_2 = \sum M_1$.

\vspace{5pt} \noindent \textbf{Processed volume $M_3$:} The volume of region removed by a clipping plane \revision{$\gamma$}{$\pi$} directly determines the efficiency of a cutting plan -- larger volume removed per cut leads to \revision{less}{fewer} times of clipping. We normalize it to $[0, 1)$ by using the volume of given model $V(\mathcal{M}\revision{}{)}$.

\vspace{5pt} \noindent \textbf{Distance to platform $M_4$:} To reflect the requirement on letting the working platform $\mathcal{P}$ always below a clipping plane \revision{$\gamma$}{$\pi$}, we define a metric as the minimal distance between \revision{$\gamma$}{$\pi$} and $\mathcal{P}$. $M_4$ is normalized by using the radius of $\mathcal{M}$'s bounding sphere. 

\begin{wrapfigure}{r}{0.4\linewidth}
  \vspace{-20pt}
  \begin{center}
   \includegraphics[width=\linewidth]{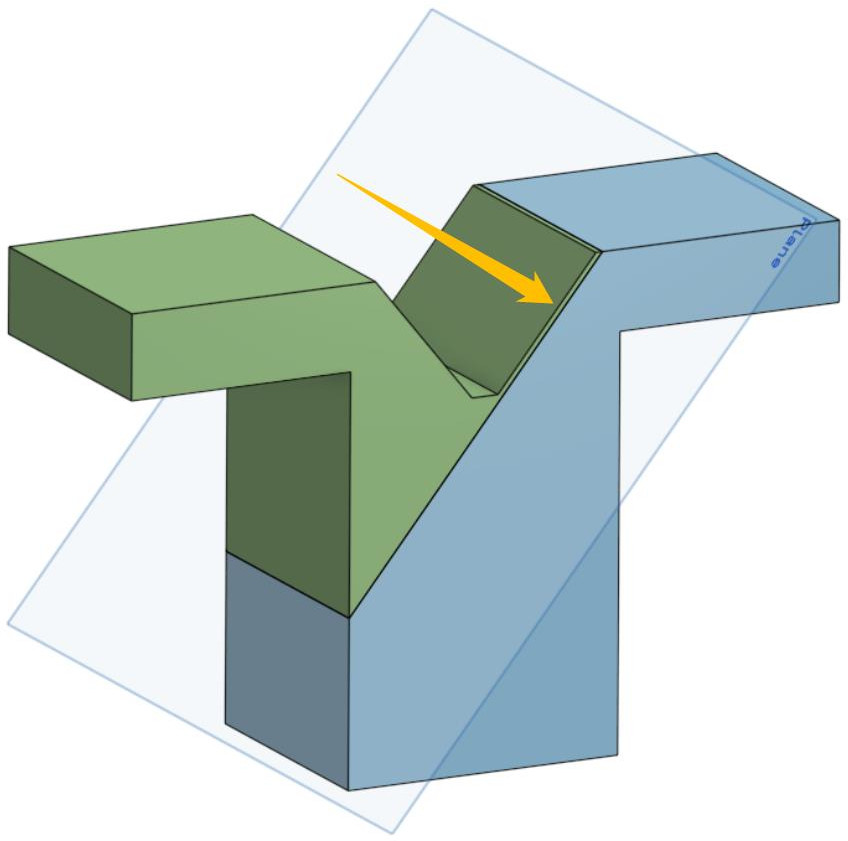}
  \end{center}
  \vspace{-10pt}
\end{wrapfigure}

\vspace{5pt} \noindent \textbf{Distance to fragile regions $M_5$:} To prevent \revision{the generation of}{from cutting through} fragile regions during \revision{3D printing}{volume decomposition}, we define the minimal distance between a clipping plane \revision{$\gamma$}{$\pi$} and all the fragile regions, which \srevision{are thin `fins' or `bridges'.} \srevision{These regions} can be detected by \revision{the method presented in}{the  geometric analysis of local curvature and feature size} \cite{luo2012chopper}. Again, this distance is normalized by the radius of $\mathcal{M}$'s bounding sphere.

\vspace{5pt} \noindent \textbf{Accumulated residual risky area $M_6$:} None of \srevision{the} above metric has considered the area that cannot be fully support-free even after decomposition -- i.e., having residual risky areas. Here we add a metric to consider the accumulated residual risky area, which is also normalized by the total risky area as $M_6 = \sum R(\mathcal{M}_k, \pi_k) / J$.

\vspace{5pt} Without loss of generality, for a candidate clipping in any stage of the planning process, it can use the vector formed by \srevision{the} above six metrics to describe its configuration. As illustrated in Fig.\ref{figTrajectory}, each candidate clipping is represented as a node during the beam-guided search. A node $n$ is denoted by $n=[M_1,M_2,M_3,\ldots,M_6]$ associated with the six metrics. In this following sub-section, we will introduce the method to select nodes kept in the beam-guided search by using the values of these metrics.

\subsection{Accelerated Search Algorithm}
Using the beam-guided search algorithm, we can obtain a list of candidate cuts with feature vectors evaluated by six metrics. The beam-guided search algorithm always keeps up to $B$ promising nodes $\mathcal{N}_k = \{n^k_1, ..., n^k_B\}$ at stage $k$.
We observe that each node $n^k_i$ may come from different parent nodes from its last stage $k-1$, and $n^k_i$ may result in different offspring nodes at the next stage $k+1$. This essentially constructs a set of trajectories starting from the input \revision{digital }{}model to the globally optimal solution of decomposition (see Fig.\ref{figTrajectory} for an example). 

When working on an input mesh $\mathcal{M}$, we can search many possible trajectories by running the beam-guided search algorithm. Each trajectory $\tau$ has a corresponding cost of $J(\tau)$. Comparing \revision{a pair of}{two} nodes $n_a^k \in \tau_A$ and $n_b^k \in \tau_B$ at the same stage $k$ belong to different trajectories \revision{$\tau_a$ and $\tau_b$}{$\tau_A$ and $\tau_B$}, we would have more preference to keep the node $n_a^k$ in the beam than $n_b^k$ when $J(\tau_A)<J(\tau_B)$ as the trajectory $\tau_A$ is more optimal. This is denoted as $n_a^k \triangleright n_b^k$. Therefore, at any stage $k$, we can always obtain \revision{ed a preference matrix $\bar{\mathbf{A}}_k=[A_{a,b}]_{B \times B}$ by letting
When a node belongs different trajectories, we always use the one with \srevision{the} smallest $J$ to update the preference matrix.}{a ranked order $\mathcal{R}_k = \{ n^k_i\}$ according to these relative relationships between the nodes.}

\begin{figure}[t]
\centering
\includegraphics[width=\linewidth]{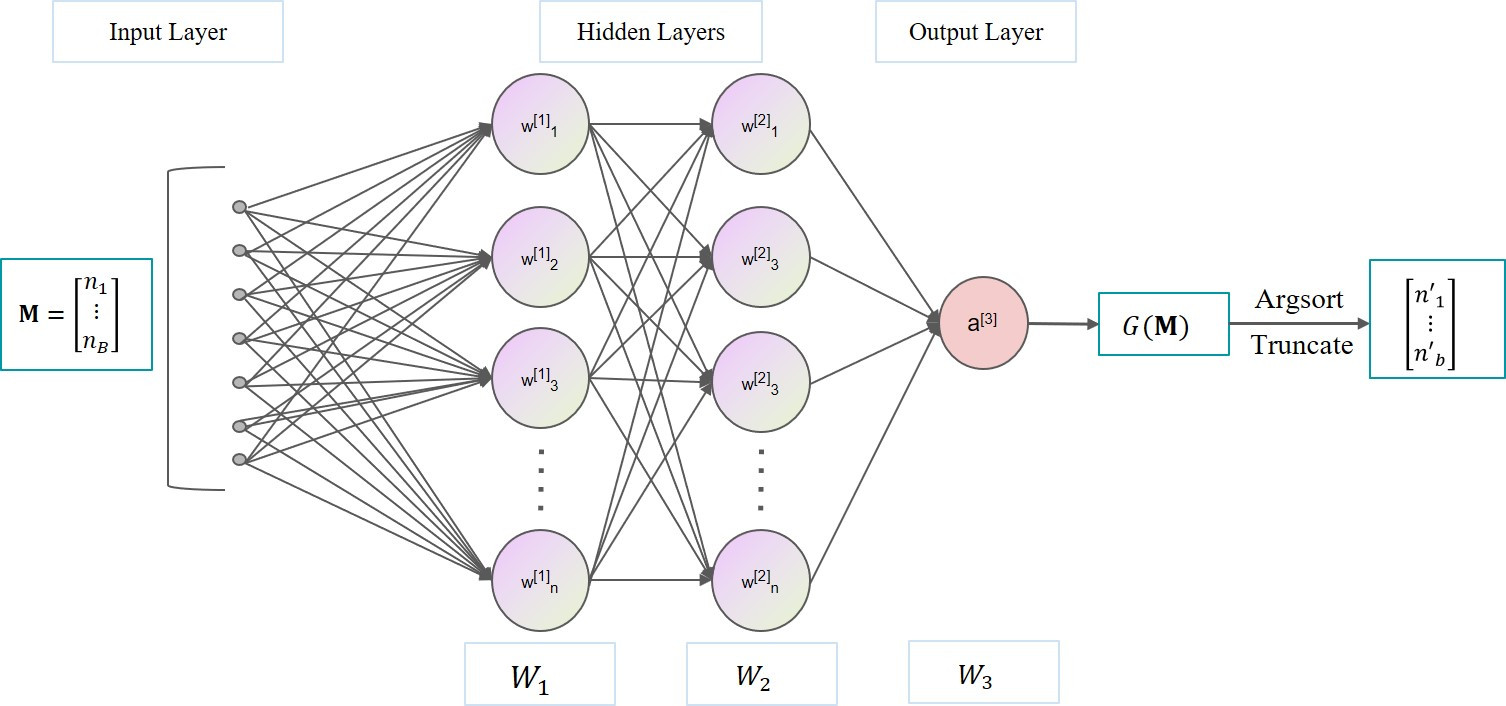}
\caption{An example that shows the pipeline of our learning to accelerate decomposition. At each stage, we use a relatively vast $B=50$ to generate candidate cuts and their metrics. Then we use the trained score function \revision{$G(n_i::n_j)$}{$G(\cdot)$} to predicate \revision{a preference matrix $A_k$ to depict relative priority between pairs}{scores} of cuts. After that, we convert the predicated \revision{$A_k$ to a voting result by row-wise summing and rank the result by arg-sorting}{scores to a ranked order by arg-sorting}. Lastly, we can select the first $b$ cuts (with $b \ll B$\revision{-- e.g., $b=6$ in this example}{}) from the selection vector for the next-round searching of decomposition.
\srevision{Note that, the input of $G(\cdot)$ are the six metrics for all the $B$ candidate cuts as a $B \times 6$ matrix $\mathbf{M}$, and the output of $G(\cdot)$ is a column of $B$ scores for these candidates.}
}\label{figPipeline}
\end{figure}

Selecting \srevision{top-}$b$ nodes from the \revision{preference matrix $\bar{\mathbf{A}}_k$ by voting}{ranked order $\mathcal{R}_k$} can have \srevision{a} high chance 
to keep nodes belong to the trajectories with \srevision{a} smaller value of $J$ in the result of selection. \revision{Here the voting is conducted by row-wise summing and then ranking the result of sum by arg-sorting.}{} In our algorithm, we are trying to learn a \revision{classifier}{scoring function $G(\cdot)$} \revision{that can generate the preference matrix $\mathbf{A}_k=[a_{i,j}]$ by using the feature-metrics such that $\mathbf{A}_k \approx \bar{\mathbf{A}}_k$. Specifically, $a_{i,j}=G(n_i::n_j)$ with the input of 12 metrics from two nodes $n_i$ and $n_j$}{from the ranked orders at different stages on different models}. With the help of the \revision{classifier $G(n)$}{scoring function $G(\cdot)$} learned from searches with large beam width. The search with smaller beam width $b$ is expected to generate results with similar quality. See also the illustration of our \revision{classifier-based voting}{scoring-function-based ranking} step for selecting $b$ nodes out of $B$ candidates given in Fig.\ref{figPipeline}. 

\subsection{\revision{Pairwise}{Listwise} Learning}
For an input mesh $\mathcal{M}$, we can obtain a collection of resultant trajectories by running the beam-guided search algorithm. Each trajectory has a corresponding cost of $J(\tau)$. Here we propose a method to convert the trajectories to \revision{pairwise}{listwise} samples to be used for learning the \revision{classifier}{scoring function} $G(\cdot)$. 
Specifically, a method is developed to sample trajectories obtained from beam-guided search with a large beam width $B=50$ on a large dataset of 3D printing models. 

Our learning method consists of four major steps. 
\begin{itemize}
\item First, we need to featurize each candidate of clipping to distinguish the differences among the other candidate. Here the six metrics introduced above in Sec.~\ref{subsec:Metrics} (i.e., $M_{1,\cdots,6}$) are used\revision{here}{}.

\item Second, we build a dataset made up of these features by running the beam-guided algorithm with a vast \srevision{value of} beam width of $B=50$. This step is very time-consuming because of the large $B$ costs more computational resources. 

\item Third, we convert the trajectories to \revision{pairwise comparison}{listwise} samples at every stage of the beam-guided search which describes the \revision{relative relationship between two different candidates of clipping}{ranking of clipping candidates}. Specifically, \revision{given two sets of trajectories $\mathcal{T}_A$ and $\mathcal{T}_B$ of the trajectories that contains $n_a^k$ and $n_b^k$ respectively, we can determine the best trajectories as $\tau_{A}=\arg \min_{\tau \in \mathcal{T}_{A}} J(\tau)$ and $\tau_{B}=\arg \min_{\tau \in \mathcal{T}_{B}} J(\tau)$. 
Then, we define}{we traverse the collection of trajectories in descending order with respect to $J(\tau)$, and use the selected node $n^k\in \tau$ to construct a set of ranked lists $\{\mathcal{R}_k\}$.}
If a node is not contained in any trajectory, it is regarded as worse than all other nodes that are contained in any trajectory. \revision{}{If a node $n^k$ was used to construct $\mathcal{R}_k$ from a trajectory $\tau_A$, it \revision{will}{would} not be used again to prevent from introducing ambiguity. We set the scores $[r_1, ..., r_b]$ of top-$b$ nodes in $\mathcal{R}_k$ as $[b, ..., 1]$ and the scores of the other nodes as zero.} The training samples are collected from all stages of \srevision{the} beam-guided search.

\item Finally, we use the \revision{pairwise}{listwise} data $\{\mathcal{R}_k\}$ to train the \revision{classifier $G(n)$}{scoring function $G(\cdot)$} by \revision{supervised learning}{learning-to-rank}. 
\end{itemize}
The resultant \revision{classifier $G(n)$}{scoring function $G(\cdot)$} will be used to evaluate every candidate of clipping in our algorithm. 

Now we have the dataset constituting of \revision{pairwise comparisons}{listwise rankings} for training. Our goal is to train a scoring system on the \revision{pairwise}{listwise} dataset to score and rank candidate cuts at each stage of \srevision{the} beam-guided search. Once the scoring system is trained, it can be utilized to replace the original 
\revision{sort-and-rank component}{selection strategy used} 
in the beam-guided algorithm. 
\revision{Our classifier is a decision-tree-based ensemble that constitutes many weak classifiers \cite{chen2016xgboost}. Detail parameters are given in the section below.}{}

\srevision{
We use uRank~\cite{urank} to train $G(\cdot)$, which formulates the purpose of ordering the nodes as selecting the most relevant ones in $|\mathcal{R}_k|$ steps. It selects all nodes that have the highest score from a candidate set at each step. To solve the classic cross-entropy issue raised by the softmax function of ratings in ListNet~\cite{cao2007learning}, it adopts multiple softmax operations and each of which targets a single node from the set of nodes that matches the ground-truth (we denote it as $c_t$, where $t$ corresponds to the step.). This method restricts the positive label appears once in candidate sets, so it only needs to select one node at each step.}

\srevision{The architecture of uRank consists of a neural network with two hidden layers with $k_1$ and $k_2$ hidden units respectively. Specifically, we have three trainable matrices $W_1 \in \mathbf{R}^{6\times k_1}$, $W_2 \in \mathbf{R}^{k_1\times k_2}$, and $W_3 \in \mathbf{R}^{k_2 \times 1}$. Let $\sigma$ be the activation function, the closed-form of $G(\mathbf{M})$ is $\sigma(\sigma(\sigma(\mathbf{M} W_1)W_2)W_3)$ with $\mathbf{M}_{B \times 6}$ being the input as six metrics of $B$ nodes. The loss function is defined as follows.}

\begin{align}
    L(G;\mathcal{R}_k)= \frac{1}{|\mathcal{R}_k|-1} \sum_{t=1}^{|\mathcal{R}_k|-1} (2^{r_t}-1) \sum_{n\in c_t} \ln{P_t(n)}
\end{align}

\noindent \srevision{where $P_t(n)$ is the likehood of selecting a node $n\in c_t$ at step $t$. The network architecture of uRank and the selection procedure are shown in Fig.\ref{figPipeline}.}

\section{Training and Evaluation}
\label{sec:results}

\subsection{Dataset Preparation and Training}
We implemented the proposed pipeline using C++ and Python, and trained the \revision{decision-tree-based ensemble using XGBoost~\cite{chen2016xgboost}}{uRank network~\cite{urank} using TensorFlow~\cite{abadi2016tensorflow}}.
The trained model and source codes are publicly accessible. The dataset collection phase is conducted on a high-performance server equipped with two Intel E5-2698 v3 CPUs and 128 GB RAM. All other tests are performed on a PC equipped with an Intel Core i7 4790 CPU, NVIDIA Geforce RTX 2060 GPU, and 24 GB RAM. We use $600$ directions sampled on the Gaussian sphere with $1$mm intervals to evaluate the metrics. The maximal self-supporting angle is set as $\alpha_{\max}=45^\circ$.

We trained our model on the \textit{Thingi10k} dataset~\cite{zhou2016thingi10k} repaired by Hu et al.~\cite{hu2018tetrahedral}. Instead of training and evaluating on the whole dataset, we extract a subset of the dataset (\revision{2061}{with 2,099} models) \revision{that satisfies}{to ensure} every model in the selected dataset should have a few risky faces that can be processed by our plane-based cutting algorithm. The training dataset for our \revision{classifier}{scoring function} is built by running the beam-guided search algorithm with $B=50$. By \srevision{the} aforementioned sampling methods, we obtain a dataset with \revision{11.96 million pairs of}{7,961 listwise} samples. We split all the dataset to 60\% samples for training, 15\% samples for validation, and 25\% data for testing. \revision{The decision-tree-based ensemble is trained by using the following parameters.}
{The numbers of hidden units we used are $k_1=100$ and $k_2=100$. In our experiments, we train the network by using maximal $1000$ epochs and a learning rate of $10^{-4}$. The early stop is invoked when no improvement is found after $200$ epochs.}

\subsection{Evaluation on Accelerated Search}
The \revision{computing time of the }{}beam-guided search algorithm\srevision{'s computing time} is significantly influenced by the chosen value of beam width $b$. \revision{We first show a comparison of computing times when using different beam widths in Fig.\ref{figBeamTimes}.}{
According to out experiments, the average computing time on $524$ test models by the conventional beam search with beam width $b=[5, 10, 20, 50]$ is at  $[2.62\times,5.38\times,9.54\times,17.77\times]$ of the average time on the beam search with $b=2$.
}

\begin{figure}[t]
\includegraphics[width=\linewidth]{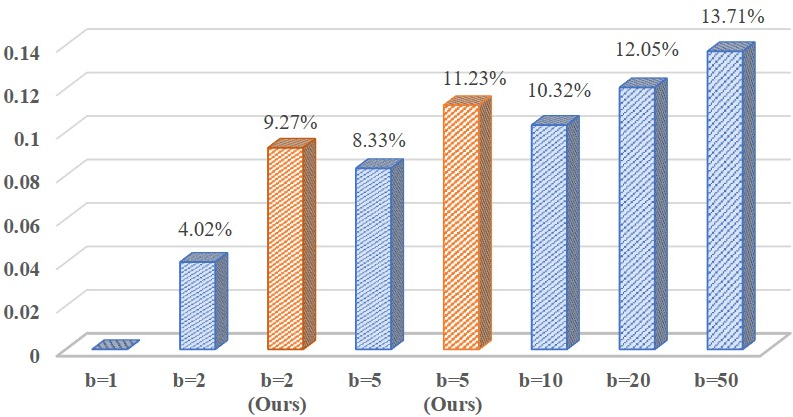}
\caption{\srevision{We use the results generated by the original beam-guided algorithm with $b=1$ as a baseline to generate comparison, where the vertical axis indicates the reduced percentage of average $J$ (i.e., Eq.(\ref{eqGlobalObjective}) on all $524$ test examples). The blue bars indicate the results of using conventional beam search, which is compared with the results of our learning-based method displayed in yellow.}}
\label{figPerformance}
\end{figure}

\begin{table}[t]
\centering
\caption{\srevision{Statistics on ranking performance}}
\begin{tabular}{|c|c|c|c|c|c|}
\hline
                        & \textbf{\begin{tabular}[c]{@{}c@{}}NDCG\\ @1\end{tabular}} & \textbf{\begin{tabular}[c]{@{}c@{}}NDCG\\ @2\end{tabular}} & \textbf{\begin{tabular}[c]{@{}c@{}}NDCG\\ @3\end{tabular}} & \textbf{\begin{tabular}[c]{@{}c@{}}NDCG\\ @4\end{tabular}} & \textbf{\begin{tabular}[c]{@{}c@{}}NDCG\\ @5\end{tabular}} \\ \hline
                        \hline
\textbf{Ours}           & \textbf{0.423}                                             & \textbf{0.455}                                             & \textbf{0.483}                                             & \textbf{0.510}                                             & \textbf{0.532}                                             \\ \hline
\textbf{RankNet}        & 0.270                                                      & 0.303                                                      & 0.335                                                      & 0.362                                                      & 0.384                                                      \\ \hline
\textbf{$\lambda$-Rank} & 0.262                                                      & 0.297                                                      & 0.326                                                      & 0.354                                                      & 0.378                                                      \\ \hline
\end{tabular}
\label{tab:ndcg}
\end{table}

After the training phase is finalized, we  \revision{now have a scoring systems to}{use the trained scoring function $G(\cdot)$ to} rank a set of features evaluated by candidate planes \revision{. We replace the simple sort-and-rank module in our beam-guided search algorithm with the classifier}{in our beam guided search,} and use $b=2$ and 5 for evaluation. To make the search procedure insensitive to minor overfitting bias, we always check if the best result ranked by the simple sort-and-rank module is in the selected beam. We run both the algorithm with the trained model and the original algorithm by different choices of $b$ on the testing dataset (\revision{511}{524} models). The statistical result in terms of \revision{performance}{improvement on the average of $J$} is \revision{shown}{given} in Fig.\ref{figPerformance}\revision{. The performance comparison (Fig.\ref{figPerformance})}{, which} shows that we can use a relatively small $b$ with the trained model to achieve a similar performance generated by a larger $b$. In other words, the search speed can be accelerated while the searched results are comparative to the ones generated using longer computing time. Meanwhile, we can improve the quality of the results produced by the original algorithm if using the trained model to select cuts.
\revision{We also observe that there are some cases (e.g., Fig.\ref{figTeaser}) where even a small $b=2$ with the trained model can produce a high-quality result, which is better than the ones generated by the original beam-guided search algorithm using $b=2$ and $5$ and the same with the ones generated by $b=20$ and $50$. }{}

\subsection{Evaluation on Ranking Performance}
\label{sec:rankingperformance}
\revision{In our learning-based acceleration algorithm, we prefer to train a network that shows better precision-at-$k$ (Prec@$k$)~\cite{mcfee2010metric}, where $k$ is the beam width we plan to deploy in the accelerated search algorithm. This measure is defined by the fraction of target items out of the first $k$ items. For example, if the actual ranking is $[1, 2, 3, 4, 5]$ and our network gives a ranking as $[1, 3, 2, 5, 4]$, the Prec@$k$ of $k=2$ is $\frac{1+0}{2}=0.5$, but the Prec@$k$ of $k=4$ is $\frac{1+1+1+0}{4}=0.75$.}{}

\revision{We use the Prec@$k$ measure to evaluate the performance of the trained network with different values of $k$ in a range of $[1, 10]$. We also use another widely used statistical measure -- mean average precision (MAP) with cut-off~\cite{baeza1999modern} to evaluate the performance, which is defined by  the averaged Prec@$k$ score of a ranking over all possible $k$.}{} 

We compare our method with other classic ranking algorithms used in information retrieval, including another listwise approach -- LambdaRank~\cite{burges2007learning} and the pairwise approach --  RankNet~\cite{burges2005learning}. We use the implementations provided in XGBoost\footnote{\url{https://github.com/dmlc/xgboost/tree/master/demo/rank}} with \revision{the same parameters used for training our scoring system}{the parameters of \{\emph{max\_depth}=8, \emph{number\_of\_boosting}=500\}}. \revision{For training the LambdaRank ($\lambda$-Rank) model, we need to specify the relevance value for each feature, thus we set the relevance values of top 5 features to $[5^2, ..., 1^2]$ and the ones of the other features to $0$.}{We use NDCG (Normalized Discounted Cumulative Gain)~\cite{jarvelin2002cumulated} to evaluate different methods.}
All experimental results are reported \srevision{using NDCG metric at position 1, 2, 3, 4, and 5 respectively} in Table \ref{tab:ndcg}. The results show that our method the best performance among all other approaches.

\subsection{\revision{Compare with other classifiers}{Feature Analysis}}

\revision{In our paper, we utilize a boosting-tree-based ensemble for pairwise learning, which is because we find out it provides the best performance. Here we use two neural-network-based classifiers, one is a classic multi-layer perceptron (MLP) consisting of two hidden layers with 24 and 6 hidden units respectively. The other is based on the same architecture but with additional recurrent hidden units to facilitate multi-stage predictions, and we set the dimension of the hidden vector as 3. Note that we train the recurrent neural network using different data pre-processing as it needs inter-stage information. Our ensemble shows the best performance among the three. As shown in~\ref{figPerformance}, our method gives $10.77\%$ and $13.48\%$ improvements on $b=2$ and $5$, while the MLP method gives $9.21\%$ and $11.87\%$ and the classifier with recurrent hidden units gives $9.34\%$ and $12.02\%$.}{}

\subsubsection{\srevision{Feature importance}}
\begin{figure}[t]
\centering
\includegraphics[width=\linewidth]{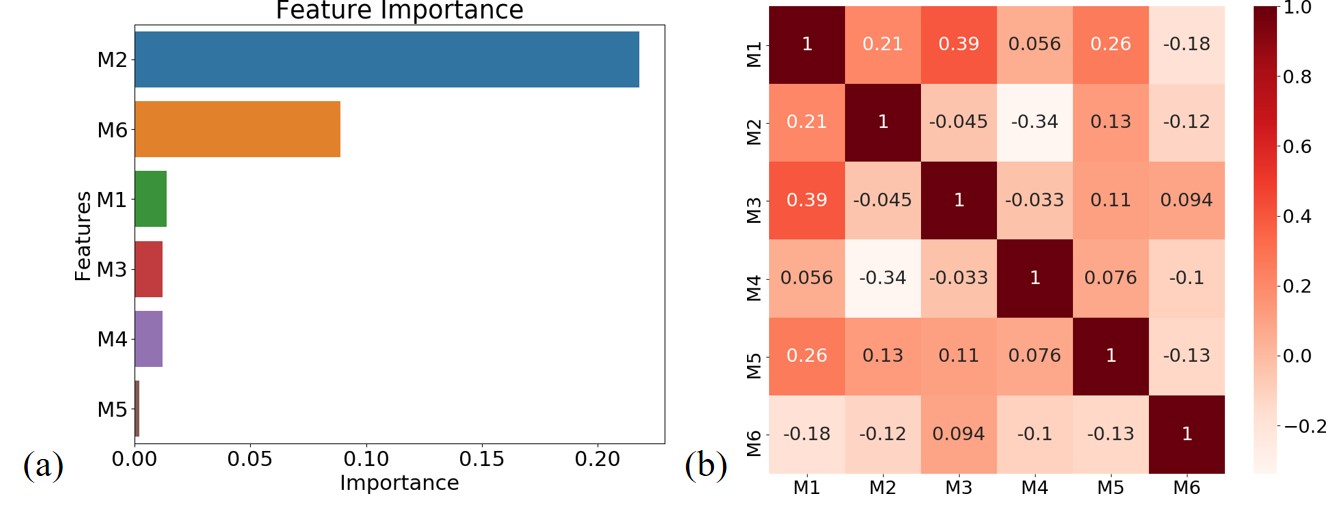}
\caption{\srevision{Feature analysis: (a) feature importance generated by permutation importance method~\cite{altmann2010permutation}, and (b) correlation analysis of six features $M_{1,\ldots,6}$ proposed in Sec.\ref{subsec:Metrics}.}}
\label{figFeatureAnalysis}
\end{figure}

\srevision{Our learning-based decomposition method extracts six features to train a neural network that can score a list of nodes. In this section, we further investigate the learned model by analyzing the features proposed in Sec.~\ref{subsec:Metrics}. We use \emph{permutation importance}~\cite{altmann2010permutation} to analyze feature importance after our model is trained. It is a widely used estimator of feature relevance in machine learning, which randomly permutes each feature column in the testing data to measure the importance of this feature. Randomization of different features will have different effects on the performance of the trained model. For any feature $M_j$, we compute its importance $I_j$ as follows.}

\begin{equation}
I_j = s - \frac{1}{K} \sum_{k=1}^{K} s_{k,j}
\end{equation}

\noindent where $s$ is the evaluation metric which could be NDCG metrics that we used in Sec.\ref{sec:rankingperformance}, and $K$ is the number of repetitions. Here we use NDCG@5 as the metric of $s$ to analyze feature importance. The experimental results are shown in Fig.\ref{figFeatureAnalysis}(a), in which $M_2$ exhibits the most importance, and $M_5$ is the least important feature among all others.

\subsubsection{\srevision{Feature correlation}}
\srevision{Correlation is a statistical measure that indicates the relationship between two or more variables. In machine learning, Pearson correlation~\cite{freedman2007statistics} is widely used to measure how the degree of the linear relationship between variables. Given two variables $M_i$ and $M_j$, their Pearson correlation $r_{M_i,M_j}$ is defined as}

\begin{equation}
    r_{M_i,M_j}=\frac{\operatorname{cov}(M_i, M_j)}{\sqrt{\operatorname{var}(M_i)} \cdot \sqrt{\operatorname{var}(M_j)}}
\end{equation}

\noindent \srevision{where var denotes the variance, and cov denotes the covariance. We use the Pearson correlation to build a correlation matrix on the sampled dataset. The heatmap visualization is shown in Fig.\ref{figFeatureAnalysis}(b). It shows that different features used in our approach are not very relevant, the most relevant feature pair is $M_1$ and $M_3$, where its $r_{M_1,M_3}=0.39$. }

\section{Conclusion and Future Work}
\label{sec:conclusion}
\begin{figure}[t]
\centering
\includegraphics[width=\linewidth]{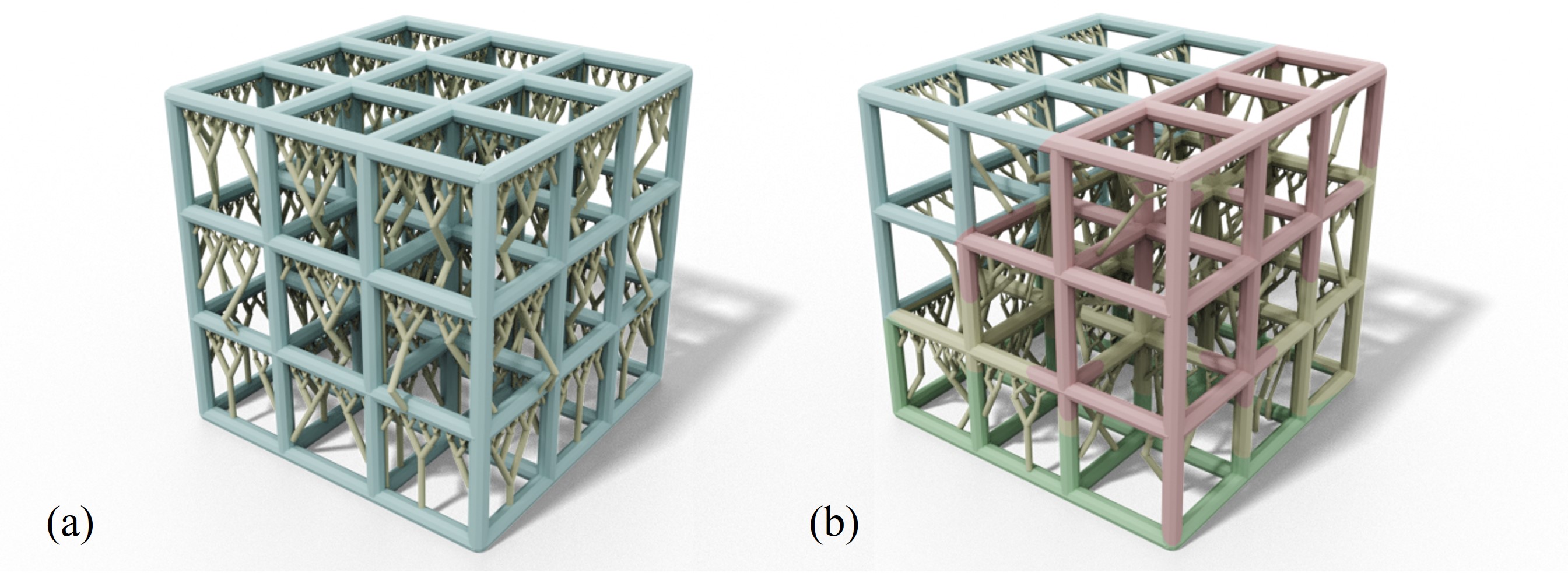}
\caption{An example of decomposing a lattice structure for multi-directional printing. (a) the cubic lattice structure with many supports. (b) the decomposed results for multi-directional printing and the generated supports.}
\label{figLattice}
\end{figure}

This paper presents an accelerated decomposition algorithm for multi-directional printing that can reduce the need of support structures. The proposed method utilizes \revision{learning-based method}{learning-to-rank techniques} to train a \revision{decision-tree-based ensemble}{neural network} that can score the candidates of clipping. We use the trained \revision{classifier to guide the search }{scoring function} to replace the simple sort-and-rank module in the beam-guided search algorithm. The computing time is reduced to \revision{1/2}{around one third} while keeping the results with similar quality. The experimental results demonstrate the effectiveness of our proposed method. We provide an easy-to-use python package and make the source code publicly accessible. In the future, we plan to investigate the capability of handling more complex objects such as periodic lattice structures. Our current method can work with models with simple lattices such as the cubic lattice structure shown in \ref{figLattice}, but the boundary of applying our multi-directional method is still unclear. Moreover, it is worth trying to study the problem of determining a given structure is suitable or not for multi-directional printing in the future.

\ifCLASSOPTIONcaptionsoff
  \newpage
\fi

\bibliographystyle{IEEEtran}%
\bibliography{ICRA20}

\begin{thebibliography}{10}
\providecommand{\url}[1]{#1}
\csname url@rmstyle\endcsname
\providecommand{\newblock}{\relax}
\providecommand{\bibinfo}[2]{#2}
\providecommand\BIBentrySTDinterwordspacing{\spaceskip=0pt\relax}
\providecommand\BIBentryALTinterwordstretchfactor{4}
\providecommand\BIBentryALTinterwordspacing{\spaceskip=\fontdimen2\font plus
\BIBentryALTinterwordstretchfactor\fontdimen3\font minus
  \fontdimen4\font\relax}
\providecommand\BIBforeignlanguage[2]{{%
\expandafter\ifx\csname l@#1\endcsname\relax
\typeout{** WARNING: IEEEtran.bst: No hyphenation pattern has been}%
\typeout{** loaded for the language `#1'. Using the pattern for}%
\typeout{** the default language instead.}%
\else
\language=\csname l@#1\endcsname
\fi
#2}}

\bibitem{Hu2015a}
K.~Hu, S.~Jin, and C.~C.~L. Wang, ``Support slimming for single material based
  additive manufacturing,'' \emph{Computer-Aided Design}, vol.~65, pp. 1--10,
  2015.

\bibitem{wu2019general}
C.~Wu, C.~Dai, G.~Fang, Y.-J. Liu, and C.~C. Wang, ``General support-effective
  decomposition for multi-directional 3-d printing,'' \emph{IEEE Transactions
  on Automation Science and Engineering}, 2019.

\bibitem{zhou2016thingi10k}
Q.~Zhou and A.~Jacobson, ``Thingi10k: A dataset of 10,000 3d-printing models,''
  \emph{arXiv preprint arXiv:1605.04797}, 2016.

\bibitem{vanek2014clever}
J.~Vanek, J.~A. Galicia, and B.~Benes, ``Clever support: Efficient support
  structure generation for digital fabrication,'' in \emph{Computer Graphics
  Forum}, vol.~33, no.~5.\hskip 1em plus 0.5em minus 0.4em\relax Wiley Online
  Library, 2014, pp. 117--125.

\bibitem{dumas2014bridging}
J.~Dumas, J.~Hergel, and S.~Lefebvre, ``Bridging the gap: Automated steady
  scaffoldings for 3d printing,'' \emph{ACM Trans. Graph.}, vol.~33, no.~4, pp.
  98:1--98:10, July 2014.

\bibitem{Hu2014}
R.~Hu, H.~Li, H.~Zhang, and D.~Cohen-Or, ``Approximate pyramidal shape
  decomposition,'' \emph{ACM Trans. Graph.}, vol.~33, no.~6, pp. 213:1--213:12,
  2014.

\bibitem{Herholz2015}
P.~Herholz, W.~Matusik, and M.~Alexa, ``Approximating free-form geometry with
  height fields for manufacturing,'' \emph{Computer Graphics Forum}, vol.~34,
  no.~2, pp. 239--251, 2015.

\bibitem{Gao2015UIST}
W.~Gao, Y.~Zhang, D.~C. Nazzetta, K.~Ramani, and R.~J. Cipra, ``{RevoMaker}:
  Enabling multi-directional and functionally-embedded {3D} printing using a
  rotational cuboidal platform,'' in \emph{Proceedings of the 28th Annual ACM
  Symposium on User Interface Software and Technology}, 2015, pp. 437--446.

\bibitem{wei18supportfree}
X.~Wei, S.~Qiu, L.~Zhu, R.~Feng, Y.~Tian, J.~Xi, and Y.~Zheng, ``Toward
  support-free {3D} printing: A skeletal approach for partitioning models,''
  \emph{IEEE Transactions on Visualization and Computer Graphics}, vol.~24,
  no.~10, pp. 2799--2812, Oct 2018.

\bibitem{muntoni2018heightblock}
A.~Muntoni, M.~Livesu, R.~Scateni, A.~Sheffer, and D.~Panozzo, ``Axis-aligned
  height-field block decomposition of {3D} shapes,'' \emph{ACM Trans. Graph.},
  2018.

\bibitem{urhal2019robot}
P.~Urhal, A.~Weightman, C.~Diver, and P.~Bartolo, ``Robot assisted additive
  manufacturing: A review,'' \emph{Robotics and Computer-Integrated
  Manufacturing}, vol.~59, pp. 335--345, 2019.

\bibitem{Keating2013}
S.~Keating and N.~Oxman, ``Compound fabrication: A multi-functional robotic
  platform for digital design and fabrication,'' \emph{Robotics and
  Computer-Integrated Manufacturing}, vol.~29, no.~6, pp. 439--448, 2013.

\bibitem{Pan2014}
Y.~Pan, C.~Zhou, Y.~Chen, and J.~Partanen, ``Multitool and multi-axis computer
  numerically controlled accumulation for fabricating conformal features on
  curved surfaces,'' \emph{{ASME} Journal of Manufacturing Science and
  Engineering}, vol. 136, no.~3, 2014.

\bibitem{peng2016fly}
H.~Peng, R.~Wu, S.~Marschner, and F.~Guimbreti{\`e}re, ``On-the-fly print:
  Incremental printing while modelling,'' in \emph{Proceedings of the 2016 CHI
  Conference on Human Factors in Computing Systems}.\hskip 1em plus 0.5em minus
  0.4em\relax ACM, 2016, pp. 887--896.

\bibitem{wu2016printing}
R.~Wu, H.~Peng, F.~Guimbreti{\`e}re, and S.~Marschner, ``Printing arbitrary
  meshes with a 5dof wireframe printer,'' \emph{ACM Trans. Graph.}, vol.~35,
  no.~4, p. 101, 2016.

\bibitem{huang2016framefab}
Y.~Huang, J.~Zhang, X.~Hu, G.~Song, Z.~Liu, L.~Yu, and L.~Liu, ``Framefab:
  robotic fabrication of frame shapes,'' \emph{ACM Trans. Graph.}, vol.~35,
  no.~6, p. 224, 2016.

\bibitem{dai2018support}
C.~Dai, C.~C.~L. Wang, C.~Wu, S.~Lefebvre, G.~Fang, and Y.-J. Liu,
  ``Support-free volume printing by multi-axis motion,'' \emph{ACM Trans.
  Graph.}, vol.~37, no.~4, pp. 134:1--134:14, July 2018.

\bibitem{shembekar2018trajectory}
A.~V. Shembekar, Y.~J. Yoon, A.~Kanyuck, and S.~K. Gupta, ``Generating robot
  trajectories for conformal three-dimensional printing using nonplanar
  layers,'' \emph{Journal of Computing and Information Science in Engineering},
  vol.~19, no.~3, p. 031011, 2019.

\bibitem{xu18supportfree}
K.~{Xu}, L.~{Chen}, and K.~{Tang}, ``Support-free layered process planning
  toward 3 + 2-axis additive manufacturing,'' \emph{IEEE Transactions on
  Automation Science and Engineering}, vol.~16, no.~2, pp. 838--850, April
  2019.

\bibitem{wu2017RoboFDM}
C.~Wu, C.~Dai, G.~Fang, Y.~J. Liu, and C.~C.~L. Wang, ``{RoboFDM}: A robotic
  system for support-free fabrication using {FDM},'' in \emph{2017 IEEE
  International Conference on Robotics and Automation (ICRA)}, May 2017, pp.
  1175--1180.

\bibitem{chen2018learning}
T.~Chen, L.~Zheng, E.~Yan, Z.~Jiang, T.~Moreau, L.~Ceze, C.~Guestrin, and
  A.~Krishnamurthy, ``Learning to optimize tensor programs,'' in \emph{Advances
  in Neural Information Processing Systems}, 2018, pp. 3389--3400.

\bibitem{adams2019learning}
A.~Adams, K.~Ma, L.~Anderson, R.~Baghdadi, T.-M. Li, M.~Gharbi, B.~Steiner,
  S.~Johnson, K.~Fatahalian, F.~Durand, \emph{et~al.}, ``Learning to optimize
  halide with tree search and random programs,'' \emph{ACM Transactions on
  Graphics (TOG)}, vol.~38, no.~4, pp. 1--12, 2019.

\bibitem{liu2009learning}
T.-Y. Liu, ``Learning to rank for information retrieval,'' \emph{Foundations
  and trends in information retrieval}, vol.~3, no.~3, pp. 225--331, 2009.

\bibitem{burges2005learning}
C.~Burges, T.~Shaked, E.~Renshaw, A.~Lazier, M.~Deeds, N.~Hamilton, and
  G.~Hullender, ``Learning to rank using gradient descent,'' in
  \emph{Proceedings of the 22nd international conference on Machine learning},
  2005, pp. 89--96.

\bibitem{burges2010ranknet}
C.~J. Burges, ``From ranknet to lambdarank to lambdamart: An overview,''
  \emph{Learning}, vol.~11, no. 23-581, p.~81, 2010.

\bibitem{cao2007learning}
Z.~Cao, T.~Qin, T.-Y. Liu, M.-F. Tsai, and H.~Li, ``Learning to rank: from
  pairwise approach to listwise approach,'' in \emph{Proceedings of the 24th
  international conference on Machine learning}, 2007, pp. 129--136.

\bibitem{guiver2009bayesian}
J.~Guiver and E.~Snelson, ``Bayesian inference for plackett-luce ranking
  models,'' in \emph{proceedings of the 26th annual international conference on
  machine learning}, 2009, pp. 377--384.

\bibitem{xia2008listwise}
F.~Xia, T.-Y. Liu, J.~Wang, W.~Zhang, and H.~Li, ``Listwise approach to
  learning to rank: theory and algorithm,'' in \emph{Proceedings of the 25th
  international conference on Machine learning}, 2008, pp. 1192--1199.

\bibitem{Lowerre1976}
B.~T. Lowerre, ``The harpy speech recognition system.'' Ph.D. dissertation,
  Carnegie Mellon University, Pittsburgh, PA, USA, 1976, aAI7619331.

\bibitem{luo2012chopper}
L.~Luo, I.~Baran, S.~Rusinkiewicz, and W.~Matusik, ``Chopper: Partitioning
  models into {3D}-printable parts,'' \emph{ACM Trans. Graph.}, vol.~31, no.~6,
  pp. 129:1--129:9, Nov. 2012.

\bibitem{urank}
X.~Zhu and D.~Klabjan, ``Listwise learning to rank by exploring unique
  ratings,'' in \emph{Proceedings of the 13th International Conference on Web
  Search and Data Mining}, ser. WSDM ’20.\hskip 1em plus 0.5em minus
  0.4em\relax New York, NY, USA: Association for Computing Machinery, 2020, p.
  798–806.

\bibitem{abadi2016tensorflow}
M.~Abadi, P.~Barham, J.~Chen, Z.~Chen, A.~Davis, J.~Dean, M.~Devin,
  S.~Ghemawat, G.~Irving, M.~Isard, \emph{et~al.}, ``Tensorflow: A system for
  large-scale machine learning,'' in \emph{12th $\{$USENIX$\}$ Symposium on
  Operating Systems Design and Implementation ($\{$OSDI$\}$ 16)}, 2016, pp.
  265--283.

\bibitem{hu2018tetrahedral}
Y.~Hu, Q.~Zhou, X.~Gao, A.~Jacobson, D.~Zorin, and D.~Panozzo, ``Tetrahedral
  meshing in the wild.'' \emph{ACM Trans. Graph.}, vol.~37, no.~4, pp. 60--1,
  2018.

\bibitem{burges2007learning}
C.~J. Burges, R.~Ragno, and Q.~V. Le, ``Learning to rank with nonsmooth cost
  functions,'' in \emph{Advances in neural information processing systems},
  2007, pp. 193--200.

\bibitem{jarvelin2002cumulated}
K.~J{\"a}rvelin and J.~Kek{\"a}l{\"a}inen, ``Cumulated gain-based evaluation of
  ir techniques,'' \emph{ACM Transactions on Information Systems (TOIS)},
  vol.~20, no.~4, pp. 422--446, 2002.

\bibitem{altmann2010permutation}
A.~Altmann, L.~Tolo{\c{s}}i, O.~Sander, and T.~Lengauer, ``Permutation
  importance: a corrected feature importance measure,'' \emph{Bioinformatics},
  vol.~26, no.~10, pp. 1340--1347, 2010.

\bibitem{freedman2007statistics}
D.~Freedman, R.~Pisani, and R.~Purves, ``Statistics (international student
  edition),'' \emph{Pisani, R. Purves, 4th edn. WW Norton \& Company, New
  York}, 2007.

\end{thebibliography}

\end{document}